\documentstyle[preprint,aps,prb]{revtex}
\begin{document}
\draft
\title{Exchange Anisotropy  in  Epitaxial and Polycrystalline NiO/NiFe 
Bilayers}

\author{R. P. Michel\cite{Remail} and A. Chaiken}
\address{Materials Science and Technology Division\\
Lawrence Livermore National Lab\\
Livermore CA 94551}

\author{C. T. Wang}
\address{Stanford University\\
Palo Alto CA 94305}

\author{L. E. Johnson}
\address{Materials Science and Technology Division\\
Lawrence Livermore National Lab\\
Livermore CA 94551}

\maketitle

\begin{abstract}

(001) oriented NiO/NiFe bilayers were grown on single crystal MgO
(001) substrates by ion beam sputtering in order to determine the
effect that the crystalline orientation of the NiO antiferromagnetic
layer has on the magnetization curve of the NiFe ferromagnetic layer.
Simple models predict no exchange anisotropy for the (001)-oriented
surface, which in its bulk termination is magnetically compensated.
Nonetheless exchange anisotropy is present in the epitaxial films,
although it is approximately half as large as in polycrystalline films
that were grown simultaneously.  Experiments show that differences in
exchange field and coercivity between polycrystalline and epitaxial
NiFe/NiO bilayers couples arise due to variations in induced surface
anisotropy and not from differences in the degree of compensation of
the terminating NiO plane.  Implications of these observations for
models of induced exchange anisotropy in NiO/NiFe bilayer couples will
be discussed.

\end{abstract}


\clearpage

\section{Introduction}

	Exchange anisotropy refers to the effect that an
antiferromagnetic (AF) layer grown in contact with a ferromagnetic
(FM) layer has on the magnetic response of the FM layer.\cite{meikel}
Exchange anisotropy is one of several magnetic interfacial
interactions, which include interlayer coupling in multilayers, that
have been intensively studied in recent years.  The most notable
changes in the FM hysteresis loop due to the surface exchange coupling
are a coercivity enhanced over the value typically observed in films
grown on a nonmagnetic substrate, and a shift in the hysteresis loop
of the ferromagnet away from the zero field axis.  The characteristics
of the AF layer and the interface between the two layers that produce
the strongest exchange bias are not well understood.  Experimental
studies and theoretical models\cite{yelon,malozemoff,mauri,soeya}
indicate that intrinsic magnetic properties of the AF such as the
magnetocrystalline anisotropy, exchange stiffness and crystalline
texture,\cite{jungblut,devasahayam,shen} as well as extrinsic
properties such as grain size, domain size and interface
roughness\cite{shen,moran} may influence the resulting response of the
FM.  Unfortunately, it is difficult to manipulate these properties
independently, or to probe the magnetic structure of the bilayer
interface directly.

Hysteresis loops of a NiO(500\AA)/NiFe(100\AA) bilayer couple measured
below and above the NiO blocking temperature, T$\rm _b$, are shown in
Figure 1 and illustrate the effects of the interface exchange
interaction.  Above the blocking temperature (T$\rm _b$=200$^{\circ}$C
$<$ the N\`{e}el temperature, T$\rm _N$ = 240$^{\circ}$C, the NiO
spins are thermally fluctuating and the NiFe film shows evidence of an
induced uniaxial anisotropy.  The NiFe film has an easy axis
coercivity of about H$\rm _{ce}$=2 Oe, and a hard axis saturation
field (not shown) of about H$\rm _s$ = 5 Oe.  After cooling to room
temperature in an external magnetic field, the NiO spins are frozen
and the interfacial magnetic interaction induces a unidirectional
anisotropy on the NiFe film which shifts the NiFe hysteresis loops
away from the zero field axis by an amount H$\rm _E$.  The direction
of the shift depends on the orientation of the NiFe layer
magnetization during field cooling.  In addition to the loop shift,
the interfacial interaction increases the coercivity dramatically.
Perpendicular to the loop shift direction, the hard axis loop (not
shown) passes nearly linearly through zero with almost no coercivity.
The 1/t$\rm _{NiFe}$ thickness dependence of H$\rm _E$ and H$\rm
_{ce}$ expected from the interfacial origin of these effects, is well
established.\cite{lin}

The dependence of H$\rm _E$ and H$\rm _{ce}$ on the NiO layer
thickness, on temperature, and on cooling field have been
documented,\cite{lin,lin2} but are not well understood at a
microscopic level.  In polycrystalline NiO films at room temperature
and with constant NiFe overlayer thickness, H$\rm _E$ and H$\rm _{ce}$
are constant for NiO layer thicknesses above about 500\AA.  As the NiO
layer thickness drops below about 350\AA, H$\rm _E$ begins to drop,
reaching zero for thicknesses below about 175\AA.  H$\rm _{ce}$
increases slightly for NiO layer thicknesses below 300\AA, peaks near
175\AA, and decreases to near zero below 100\AA.  The decrease in
H$\rm _E$ in bilayers with thin NiO thickness is not accounted for by
a reduction in the N\'{e}el temperature due to finite size
effects\cite{borchers} and instead indicates a length scale associated
with the frozen AF spin configuration.

For NiO layers thicker than 400\AA, H$\rm _E$ drops with increasing
temperature, reaching zero at about 200$^{\circ}$C, which is called
the blocking temperature T$\rm _b$.  When the NiO layers are thinner
than 400\AA, the blocking temperature is reduced.\cite{lin2} T$\rm _b$
is thought to be associated with a thermal activation energy for the
domain configuration in the NiO.  Experiments have shown that the
decrease in H$\rm _E$ with temperature can be accounted for by
assuming a distribution of blocking temperatures,\cite{soeya} possibly
indicating a distribution of domain sizes or anisotropy energies.

Typically NiFe is deposited on top of NiO to form a NiFe/NiO exchange
couple.  A bias field (20-200 Oe) is applied during deposition to
induce a uniaxial anisotropy in the NiFe layer.  It is thought that
the interaction of the aligned NiFe spins at the interface with the
NiO during deposition influences the AF spins in the NiO since the
applied bias field is too weak to induce ordering in the NiO spins
directly.  In turn the NiO spin arrangement at the interface induces a
unidirectional surface anisotropy in the NiFe.  Heating bilayers above
T$\rm _b$ and cooling in a field has been reported to
increase\cite{lin2} and to decrease\cite{lin,carey} H$\rm _E$ relative
to the as-deposited values.  Whether the magnitude of the bias field
or the cooling field strongly influences H$\rm _E$ has not been well
established.\cite{moran2}

Exchange couples which incorporate FeMn, NiMn, PdMn, IrMn, Pd-Pt-Mn,
NiO, and NiCoO antiferromagnetic layers are currently under study for
use in magnetoresistive sensors and magnetoresistive and spin-valve
based hard disk readback heads.\cite{kanai} The exchange anisotropy is
employed to achieve the optimum sensitivity bias configuration in the
sensor and to reduce noise by stabilizing
domains.\cite{lin,hamakawa,ciureanu} In this paper we focus on the
oxide AF materials which share the same rocksalt crystal
structure. The AF spin configurations and exchange coupling properties
of the Mn-based materials are significantly different from the oxide
materials and thus must be considered separately.  The oxide films
proposed for applications are polycrystalline with relatively small
grain sizes.  Achieving a clearer understanding of how
magnetocrystalline anisotropy and texture influence the exchange
anisotropy, however, requires that films with a high degree of
crystalline perfection be examined as well.

The NiO spin structure is relatively simple, however the large number
of domain configurations and domain walls in a multidomain sample make
theoretical models of exchange anisotropy in NiO/NiFe bilayers
considerably more challenging.\cite{farztdinov,roth,roth2,sievers} NiO
has a cubic FCC NaCl crystal structure above its N\'{e}el temperature.
Below the N\'{e}el temperature there is a slight distortion of the NiO
lattice in a $\rm \left< 111 \right>$ direction ($ \Delta \ell/\ell
\approx 4.5x10^{-3}$).\cite{slack} A strong negative uniaxial
anisotropy accompanies the contraction, resulting in an easy plane
defined by K$\rm _1 \approx 1x10^{6}$ erg/cm$^3$.\cite{sievers} Sheets
of ferromagnetically aligned spins form in the (111) planes defined by
the contraction axis,\cite{roth} with the Ni spins in neighboring
sheets oppositely aligned. Within a (111) plane the direction of the
spin axis is determined by a second 3-fold anisotropy (K$\rm _3$) that
is roughly three orders of magnitude weaker than K$\rm
_1$.\cite{slack,mcguire}

The AF domain configurations in NiO have been studied both
experimentally and theoretically.  There are four possible (111)
directions in a NiO crystal from which the contraction axis may
choose, and 3 spin directions once the contraction axis is defined.
Thus there are 4x3=12 distinct possible AF domain configurations in
NiO below T$\rm _N$.  Since the four (111) directions in the cubic NiO
are nominally equivalent, local inhomogeneities determine which (111)
axis becomes the contraction axis in different regions of the crystal.
Applied magnetic fields and strain can make one (111) direction more
kinetically favorable, and thereby influence the distribution of AF
domains.  The magnetic susceptibility of the NiO is largest parallel
to the contraction axis, and so this axis tends to align parallel to
strong applied fields.\cite{roth3} It has proven to be experimentally
nearly impossible, however, to create a macroscopic NiO specimen with
a single contraction axis by cooling in a magnetic field.  Once the
sample temperature has been lowered below the N\'{e}el temperature,
domain walls become strongly pinned\cite{roth2,roth3,saito} and
extremely large fields are required to change the AF domain
configuration.

In this study, we compare the magnetic properties of polycrystalline
and epitaxial (001) NiO/NiFe bilayers deposited simultaneously.  We
also compare epitaxial (001) NiO/NiFe bilayers with the deposition
bias field, H$\rm _b$, aligned along different in-plane NiO
crystalline axes.  The results of these studies are interpreted in
terms of induced anisotropies at the NiO/NiFe interface.

\section{Experimental Methods}

	The single films and bilayers were grown in a multilayer
deposition system using ion beam sputtering (IBS). The system has been
described in detail elsewhere.\cite{fesimmm,fesiprb} Single crystal
polished (001) oriented MgO substrates and Si substrates with native
oxide layers were placed side by side in substrate holders with bias
magnets. The bias magnets produce a uniform magnetic field, H$\rm _b$,
of about 300 Oe at the substrate surface.  The substrate temperature
was monitored but not controlled and reached about 80$^{\circ}$C
during deposition.  The NiO layers were grown using a new IBS
process.\cite{intermagnio} Briefly, this new process is simpler than
the more widely used reactive sputtering technique in that an oxygen
partial pressure is not required during NiO deposition.  The NiO is
deposited by directly sputtering a NiO target with a neutralized 750V,
30 mA Ar-ion beam which produces a deposition rate of about 0.2
\AA/sec.  The metal NiFe layers were then grown immediately on top of
the NiO using a 500V, 20 mA ion beam without neutralization.  The Ar
gas pressure during deposition was about 0.25 mTorr. The Ni:O ratio of
films produced using this process was measured using Rutherford
back-scattering and was determined to be 1:1 to within 1\%.  NiFe
films were deposited from a Fe$\rm _{19}Ni_{81}$ sputter target.

	The morphology of the NiO/NiFe bilayers was probed using x-ray
diffraction (XRD).  Symmetric x-ray scans were performed on a 18 kW
Rigaku rotating anode diffractometer with a scattered beam
monochromator using Cu K$_{\alpha}$ radiation.  Phi scans were
performed on a four-circle goniometer using Cu radiation at Stanford
University.

	The magnetic properties of the bilayer films were measured
with a vibrating sample magnetometer (VSM) equipped with two sets of
orthogonal pick-up coils.  The use of two sets of pick-up coils allows
the simultaneous measurement of both the longitudinal and transverse
magnetization curves, as is often done with Kerr
magnetometry.\cite{florczak} The magnetization curve of a 500 \AA\
thick NiFe film grown on MgO (001) and Si (not shown) shows that H$\rm
_b$ applied during deposition induces a uniaxial anisotropy in the
NiFe.  The value of the uniaxial anisotropy is determined from the
hard axis saturation field, H$\rm _s$, to be K = 2000 erg/cm$^3$ given
that H$\rm _s$ = 2K/M$\rm _s$ in the polycrystalline NiFe film. In
addition to K we also observe a four-fold magnetocrystalline
anisotropy in the epitaxial NiFe film of about K$\rm _1$ = -500
erg/cm$\rm ^3$ (H$\rm _s$ = 2(K+K$\rm _1$)/M$\rm _s$ for a $\rm \left<
110 \right>$ hard axis in the (001) plane).  These values are
consistent with those expected for NiFe films.\cite{chikazumi}

\section{Experimental Results}

	A comparison of the XRD spectra for NiO/NiFe bilayers grown
simultaneously on MgO (001) and oxidized silicon is shown on Figure 2.
The films deposited on oxidized Si wafers are polycrystalline as shown
by the presence of (111), (002) and (022) NiO Bragg peaks.  The
average grain size calculated using the Scherrer formula from the full
width at half maximum (FWHM) of the peaks is 100\AA\ to 200\AA.
Because the NiO and MgO crystal structures are nearly identical with
only slightly different lattice parameters, (MgO: a=4.213\AA, NiO:
a=4.177\AA\ or 0.9\% difference) the NiO (002) Bragg peak of the film
on MgO is obscured under the strong substrate peak.  No NiO (111) or
(022) Bragg peak intensity was observed in the XRD spectra of the
bilayer grown on MgO, however.  Instead a strong (002) peak from the
NiFe deposited on the NiO was present with a correlation length
limited by the thickness of the film (100\AA) and a rocking curve
width of 1$^{\circ}$-2$^{\circ}$ FWHM.  Phi scans at the NiFe (011)
peak position show that the NiFe layer grown on top of the NiO layer
is epitaxially oriented relative to the MgO substrate (Figure 2b). The
epitaxy of the NiFe shows that the intermediate NiO layer is also
oriented in-plane with respect to the MgO substrate.  As discussed in
the previous section, NiFe films grown directly on MgO (001) were also
found to be epitaxial.  Comparison of Kiessig fringes in the low-angle
symmetric XRD spectra (not shown) indicate that the interfaces of the
epitaxial bilayer are rougher (8-12 \AA) than the polycrystalline
bilayer (2-3 \AA).

	The hysteresis loops of polycrystalline and epitaxial NiO
(500\AA)/NiFe (100\AA) bilayers grown simultaneously are shown in
Figure 3.  In the epitaxial film the bias field during deposition
H$\rm _b$ was applied along a MgO [100] axis.  Hysteresis loops
parallel and perpendicular to H$\rm _b$ are shown.  The
polycrystalline bilayer (Fig. 3a) illustrates the usual exchange
anisotropy behavior: there is a shift in the easy axis hysteresis loop
of H$\rm _E$ = 52 Oe and an increase in the NiFe coercivity from its
free value of about H$\rm _{ce}$ = 2 Oe to H$\rm _{ce}$ = 30 Oe.  The
hard axis loop shows almost zero coercivity and saturates at about
2H$\rm _E$.  The loop parallel to H$\rm _b$ for the epitaxial bilayer
(Fig. 3b) shows a shift of H$\rm _E$ = 20 Oe and a coercivity of H$\rm
_{ce}$ = 26 Oe.  The shape of the hard axis magnetization
(perpendicular to H$\rm _b$ ) in the epitaxial film is qualitatively
different from the nearly linear hard axis loop observed in the
polycrystalline films.  Figure 3c shows transverse magnetization data
for the polycrystalline and epitaxial films with H $\perp$ H$\rm _b$,
and indicates the nonlinearity in the epitaxial bilayer's hard axis
loop is due to a nearly 90$^{\circ}$ reorientation of the
magnetization vector.

	In figure 4 another set of magnetization data for
polycrystalline and epitaxial NiO/NiFe bilayers grown simultaneously
is shown but now with H$\rm _b$, the bias field applied during
deposition, aligned with an in-plane MgO [110] axis.  The
polycrystalline bilayer (Fig. 4a) has an exchange field of H$\rm _E$ =
66 Oe and an easy-axis coercivity of H$\rm _{ce}$ = 34 Oe.  The hard
axis loop once again saturates at about H$\rm _s$ = 2H$\rm _E$ and has
coercivity less than 1 Oe.  The epitaxial bilayer (Fig. 4b) has H$\rm
_E$ = 36 Oe, and H$\rm _{ce}$ = 42 Oe.  The hard axis magnetization
data shown in figure 4b and 4c shows similar behavior to that seen in
figure 3b and 3c.  The similarity between figure 3b and 4b reveals
that the nonlinearity observed in the hard axis magnetization curves
of the epitaxial films is induced by H$\rm _b$ and is not referenced
to the underlying crystal structure of the NiO.  The variation in
H$\rm _E$ for the polycrystalline samples shown in figures 3a and 4a
results from uncontrolled variations in the deposition conditions and
serves as a measure of the run-to-run reproducibility of the growth.

	We have grown epitaxial bilayers in reverse order to better
understand why this configuration typically shows lower exchange
anisotropy than do bilayers with NiO on the bottom.\cite{ambrose}
Figure 5 shows easy axis magnetization loops for
NiFe 100\AA/NiO 500\AA\ bilayers grown simultaneously on oxidized
silicon and MgO (001) substrates.  In this configuration the choice of
substrate has much less effect on the hysteresis loop. The exchange
shift is H$\rm _E$ =13 Oe and H$\rm _{ce}$ = 5 Oe in both films.  XRD
shows that both NiO films are polycrystalline.  These data demonstrate
that the growth mode of NiO on NiFe is significantly different than
that of NiFe on NiO, which explain the difference in the exchange
anisotropy observed in the two configurations.  Recently, however,
large bias fields (0.04 erg/cm$\rm ^2$) were reported in ``top'' spin
valves grown with the NiO layer on top of a NiFe layer using reactive
RF sputtering.\cite{everitt}

\section{Analysis}

	In order to model the field dependence of the NiFe
magnetization, we start with the simplest energy equation that
contains only a unidirectional anisotropy term and a Zeeman term
describing interaction with the external field.  The energy equation
takes the form:

\begin{equation}
E/M \, = \, -H \cos(\theta) \; - \; H_E \cos(\theta \: - \phi)
\end{equation}
                 
\noindent where H$\rm _E$ is the effective unidirectional anisotropy
field, $\phi$ is the angle between the bias field H$\rm _b$ and the
applied field H, and $\theta$ is the angle between H and the
magnetization.\cite{meikel} (This form ignores the induced uniaxial
anisotropy in the NiFe layer, which is small compared to H$\rm _E$.)
Assuming the magnetization reorients by rotation following the minimum
energy solution, the hard axis magnetization is:
     
\begin{equation}
\frac{M(H)}{M_s} \, = \, \frac{H}{\sqrt{ ( H^2 \: + \: H_E ^2 )} }
\end{equation}

\noindent Under these assumptions, the easy axis magnetization should
have zero coercivity and change sign at H = H$\rm _E$, and the hard
axis magnetization should approach saturation asymptotically. The best
fit of equation 2 to the measured hard axis magnetization for the
polycrystalline film in figure 3a predicts H$\rm _E$ = 83 Oe which is
inconsistent with the measured easy axis value of H$\rm _E$ = 52 Oe.
By increasing the size of uniaxial anisotropy above the usual value
for soft NiFe alloys, we can consistently fit the easy and hard axis
behavior observed in figure 3a, and 4a, and qualitatively account for
the easy-axis coercivity.

\begin{equation}
E/M \, = \, -H \cos ( \theta  ) \; - \; H_E \cos ( \theta \: - \:
\phi ) \; + \; H_K \cos ^2 ( \theta \: - \: \phi )
\end{equation}

The predicted analytical form of M(H) is complicated.  However, we can
consistently fit the easy and hard axis behavior with H$\rm _E$ = 53
Oe and H$\rm _K$ = 30 Oe, given that equation 3 predicts the
magnetization approaches saturation with $\rm H_s \: \approx \: 2(H_E
\: + H_K)$.  The uniaxial term is significantly larger than that
observed in single films of NiFe (H$\rm _K$ of NiFe = 5
Oe).\cite{chikazumi} The increase of H$\rm _K$ comes from the
interfacial interaction with the NiO.  Strain induced at the NiO/NiFe
interface may be a source of uniaxial anisotropy, however, the small
magnetostriction of the permalloy, combined with the small tetragonal
distortion of the NiO below its N\'{e}el temperature make this an
unlikely explanation for the large uniaxial anisotropy observed here.

	The large uniaxial term in the energy equation needed to
consistently model the easy and hard axis data helps to account for
the coercivity in the easy axis loop.  It is well known that, in the
presence of a uniaxial term, the energy equation contains local energy
minima in addition to global minima for a range of applied
fields.\cite{dieny,dieny2} Local energy minima can pin the
magnetization and temporarily delay the obtainment of the absolute
energy minimum configuration.  Since the magnetization loops show that
NiFe moment reverses by rotation and follows the local energy minimum,
the easy axis coercivity predicted by equation 3 is H$\rm _{ce}$ =
H$\rm _K$ = 30 Oe.  In single NiFe films, however, H$\rm _{ce} <$ H$\rm
_K$ indicating the reversal occurs through domain wall motion rather
than rotation.  The easy axis energy surfaces predicted by equation 1
do not contain local minima, however, those by equation 3 do.  Thus
the static interfacial interaction at the NiO/NiFe surface anisotropy
can account for much of the coercivity observed in the easy axis loop.

	It is interesting to note that the ratio of the H$\rm _E$ to
H$\rm _{ce}$ for a wide range of IBS polycrystalline NiO/NiFe bilayers
appears to have a characteristic maximum value.  Figure 6 shows H$\rm
_E$ plotted vs. H$\rm _{ce}$.  The dotted line is a guide to the eye
showing H$\rm _E$ = 1.8 H$\rm _{ce}$ + 14 Oe.  Fig. 6 implies that
increases in H$\rm _E$ and H$\rm _{ce}$ come together in the best
polycrystalline NiO/NiFe bilayers.  The data in figure 6 imply that
H$\rm _E$ and H$\rm _K$ in equation 3 do not vary independently in
NiO/NiFe bilayers.  On the other hand, films with small H$\rm _E$ and
large H$\rm _{ce}$ occur since there are many sources of coercivity in
thin NiFe films, many not directly related to the surface exchange
interaction.  The highest H$\rm _E$ to H$\rm _{ce}$ ratio in NiO/NiFe
bilayers published in the literature,\cite{lin,lin2} deposited using
reactive sputtering is approximately 2.2, which is similar to the
value observed in the IBS films.

The slope of the line in figure 6 depends on the intrinsic
anisotropies present in NiO.  The H$\rm _E$/H$\rm _{ce}$ ratios we
observe in coupled IBS NiFe/NiCoO bilayers typically lie above this
line.  Further, a typical H$\rm _E$/H$\rm _{ce}$ ratio for NiFe/FeMn
bilayers is 25.\cite{kim} The higher ratio observed in general in
Mn-based AF exchange couples may be due to the higher
magnetocrystalline anisotropy or the reduced symmetry of the Mn-based
antiferromagnets.\cite{jungblut,lin} These differences produce a
interface anisotropy that is closer to pure unidirectional in
FeMn/NiFe bilayers compared to the unidirectional plus uniaxial
anisotropy found in NiO/NiFe bilayers.

	Turning now to the magnetization observed in the epitaxial
bilayers in figures 3b, and 4b, the shape of the hard-axis
magnetization curves can be predicted by adding a cubic anisotropy to
the energy equation (1):

\begin{equation}
H_{k1} sin ^2 ( \theta \: - \: \phi ) \cos ^2 ( \theta \: - \: \phi )
\end{equation}

\noindent The data in figure 3b are reasonably well reproduced with
H$\rm _E$ = 20 Oe , H$\rm _K$ = 30 Oe.  The cubic anisotropy produces
an energy minimum perpendicular to the unidirectional anisotropy (and
to H$\rm _b$) which qualitatively changes the hard axis loop shape.

	In addition, the presence of a cubic anisotropy produces local
energy minima in the energy surface describing the bilayer
magnetization reversal.  As discussed previously, these minima can be
used to qualitatively account for the coercivity observed in the hard
axis magnetization loop.  Qualitatively, as H decreases from a large
positive value, the NiFe layer moment at first remains in a local
energy minimum parallel to H, and then shifts suddenly from that
minimum to the energy minimum derived from the unidirectional
anisotropy term, perpendicular to H and parallel to H$\rm _b$.
Transverse magnetization data (Fig. 3c, 4c) for H $\perp$ H$\rm _b$
reinforce this description.  As the longitudinal magnetization (M$\rm
_x$) decreases, the transverse magnetization (M$\rm _y$) increases
abruptly and reaches a plateau as the NiFe layer moment settles into
the global energy minimum perpendicular to H.  This is in contrast to
the transverse hard-axis behavior of the polycrystalline bilayer
couple which shows a smooth rotation of the NiFe layer moment and no
plateau.

	Calculated magnetization curves that qualitatively reproduce
the experimental data for the epitaxial bilayers are shown in figure
7a,b.  The curves were calculated using an energy equation with a
unidirectional and a cubic anisotropy:

\begin{equation}
E/(M*H_E) \, = \, -H/H_E\cos(\theta) \; - \; \cos(\theta \: - \: \phi)
\; + \; H_{K1}/H_E \cos^2 (\theta \: - \: \phi) \sin^2 (\theta \: - \:
\phi)
\end{equation}

\noindent where H$\rm _{K1}$/H$\rm _E$ = 1.5.  The magnetization in
Figure 7a is assumed to reverse by rotation and to find the absolute
minimum energy configuration.  The calculation reproduces the steps
observed in the epitaxial hard axis loops.  As in the case of the
polycrystalline bilayers, if we assume the vector magnetization sticks
in local energy minima and only achieves the absolute minimum when its
path is unobstructed by an energy barrier, we can qualitatively
account for the coercivity observed in the easy and hard axis loops as
shown in figure 7b.

	From the similarity of the hard axis loops in figure 3b,c and
4b,c, where the bias field is applied along a different NiO crystal
axis, it is clear that the cubic term is induced by the bias field
applied during deposition, and is not referenced to the NiO or the
NiFe crystal axes.  Thus the data are not consistent with a
magneto-crystalline anisotropy in the NiFe or the NiO.  In contrast,
epitaxial NiFe films deposited directly on MgO (001) show induced bulk
uniaxial and cubic magnetocrystalline anisotropy terms which are
nearly an order of magnitude smaller than those needed to describe the
NiO/NiFe loops. The cubic anisotropy must arise from the same
interfacial interaction with the NiO that produces the exchange
anisotropy.  Ferromagnetic resonance or Brillouin light-scattering
measurements on these bilayer films may give a more quantitative
determination of the anisotropy values.

	The above analysis assumes the NiFe/NiO interfacial anisotropy
is unmodified by rotation of the NiFe moment during magnetization
reversal.  This assumption is strictly false.  The observation of a
training effect and the presence of rotational hysteresis in NiFe/NiO
bilayers even at very high applied fields\cite{paccard} clearly show
that NiO spin dynamics are present during the NiFe reversal.  However,
a static induced surface anisotropy does describe many of the main
features of the NiFe/NiO magnetization curves.  The contributions of
irreversible NiO spin dynamics on the NiFe loop are second-order
effects.  More experimental work is needed to measure and understand
the NiO spin dynamics during the NiFe magnetization reversal process.

	Once again it is interesting to compare the behavior of
NiFe/NiO and NiFe/FeMn exchange couples.  As with NiO/NiFe exchange
couples, there is a strong deposition order dependence in NiFe/FeMn
exchange couples.  However, in this case it is the NiFe that should be
deposited first in order to achieve a large exchange
bias.\cite{stoecklein} The order dependence of NiFe/FeMn exchange bias
has been found to arise from changes in growth mode when the order of
deposition is reversed.  The (111) textured NiFe surface serves as a
template for the antiferromagnetic $\gamma$ phase of FeMn.  In the
absence of the NiFe template the FeMn does not achieve the $\gamma$
phase and instead forms in the nonmagnetic $\alpha$ phase, and no
exchange bias is observed.  In experiments where the $\gamma$ FeMn is
stabilized through epitaxy with a single crystal substrate, exchange
bias is observed in NiFe deposited on top.\cite{jungblut} Further,
when the $\gamma$ FeMn was grown in different crystalline
orientations, exchange bias in the NiFe grown on top was observed in
every case.\cite{jungblut} The ratio of H$\rm _E$ to H$\rm _{ce}$ for
the FeMn/NiFe bilayers was different for the different crystalline
orientations, however.\cite{jungblut} Changing the order of NiO/NiFe
bilayer deposition does not change the crystalline phases of the
individual layers, but instead produces more subtle changes at the
interface that lead to differences in the exchange anisotropy.

\section{Discussion}

	There are two requirements for achieving shifts in the
hysteresis loop of a ferromagnetic (FM) film deposited on an
antiferromagnetic (AF) film and these requirements are sometimes at
odds.  First, there must be an uncompensated interaction at the
interface.  Second, spins in the AF layer must remain pinned as the FM
film undergoes a reversal.  The most obvious way to produce a
magnetically uncompensated surface in the NiO is to terminate the NiO
layer on the (111) face normal to the contraction axis.  These (111)
planes contain sheets of aligned spins.  The anisotropy within this
(111) plane is weak, however, and so the NiO spins may not be strongly
pinned and may rotate within this plane during a NiFe reversal.  In
addition, the presence of domain walls in the NiO layer perpendicular
to the interface will increases the degree of compensation due to
averaging of the spin orientations over the surface.\cite{malozemoff2}
The presence of interface roughness makes it still more difficult to
predict if the surface interaction will be
uncompensated.\cite{malozemoff} The in-plane anisotropy is strongest
when the contraction axis is forced into the plane of the film, but in
this orientation the bulk NiO spin configuration predicts the surface
will be compensated even at very short length scales.  In
polycrystalline NiO films, NiO spin rotations and domain wall dynamics
are strongly influenced by grain boundaries and crystalline defects,
in addition to the intrinsic magnetocrystalline
anisotropy.\cite{roth2,roth4,saito}

	Our results clearly show that the crystalline structure of the
NiO in NiO/NiFe thin film exchange couples does not strongly influence
the spin structures that are responsible for the uncompensated
interaction.  Instead, it is the influence of the applied field
indirectly through alignment of the NiFe layer and the subsequent
NiFe/NiO interfacial interaction which determine the spin distribution
at the NiO/NiFe interface.  Coupling at the interface between the NiO
and the aligned NiFe layer must force the Ni spins in the NiO on
average to be collinear, parallel to H$\rm _b$.  These interactions
force the spins at the surface of the NiO layer to distort producing
an uncompensated interface nearly independent of the crystalline
orientation or morphology at the interface.  In other words, the
interfacial NiO spin configuration is not the same as the bulk NiO
spin configuration.  The distorted spin configuration at the NiO
surface must also be strongly coupled to the bulk of the NiO layer,
anchoring the surface spins, so they do not significantly reorient
during subsequent ferromagnetic reversals once the configuration is
frozen in.  This picture is consistent with the one presented by
Schlenker\cite{schlenker} and Stoecklein\cite{stoecklein} who suggest
the frustrated interactions at the AF/F interface may be similar to
those in a spin glass.

	It is not consistent to assume that the formation of an
aligned NiFe layer at the NiO surface during deposition can distort
the NiO interfacial spins to create an uncompensated interaction, but
yet the distorted configuration remains frozen during subsequent post
deposition NiFe reversals.  The time scale for the NiO spins dynamics
must therefore be shorter during deposition than during routine room
temperature magnetic measurements. It follows that the temperature
during deposition is elevated close enough to T$\rm _N$ to allow the
distortion of the surface NiO spin configuration during the short time
scale of a deposition. Thus one could predict that NiFe deposition
onto a cooled NiO thin film would produce no loop shift.  In order to
address this hypothesis, we have deposited IBS NiO/NiFe bilayers on
substrates clamped to a thick Cu plate that is cooled to
-120$^{\circ}$C.  Large loop shifts were still observed in these
films.  The distorted spin region described in the previous paragraph
that creates the uncompensated NiO surface may be relatively thin, and
thus only the temperature at the surface of the growing film need
approach T$\rm _N$ in order for a loop shift to be observed in the as
deposited films.  It is likely that even when the substrates are
cooled from the back side, energy transported to the film surface by
deposited material raises the surface temperature above T$\rm _N$.  A
more effective test of the above hypothesis would employ a low energy
NiFe deposition technique such as evaporation in conjunction with
substrate cooling.  Stoecklein suggests that simply the presence of
the AF/F interface produces the spin frustration needed to produce a
loop shift,\cite{stoecklein} however this model cannot explain the
persistence of the loop shift during post deposition NiFe reversals.

  	It is easy to see that in as-deposited bilayers in which
effectively only a thin surface layer has been field-cooled through
T$\rm _N$ , we produce a different distorted NiO spin configuration
than in field-annealed bilayers.  In this way we can qualitatively
account for changes in as-deposited and field-annealed H$\rm _E$.
During reverse-order deposition (NiFe first), the thermal development
of the NiO surface layer will be different than during forward-order
(NiO first).  One can speculate that the presence of the thick metal
NiFe layer will more efficiently cool the interface during
reverse-order deposition, resulting in a thinner distorted layer and a
smaller H$\rm _E$.  In analogy with spin glass models, the NiO/NiFe
interface may possess a large number of nearly equal energy
configurations, each producing a unique value of H$\rm _E$.  In spin
glasses, configurations are separated by large energy barriers
confining the system to a small region in phase space for short time
scales.  Changing the thermal and magnetic history of the interface,
allows the spin configuration to quickly explore different local
minima.\cite{weissman}

	Clearly exchange couples based on polycrystalline rather than
epitaxial NiO layers show the most promise for applications.  In
polycrystalline exchange couples the increased disorder of the NiO
spins leads to larger exchange bias and smaller easy axis coercivity
than in epitaxial films.  The larger interface roughness in the
epitaxial bilayers relative to the polycrystalline ones may account
for the observed differences in the magnitude of the exchange
field.\cite{shen} The higher degree of crystalline perfection in the
epitaxial bilayers leads to more complicated induced surface
anisotropy.  Stronger pinning in polycrystalline films may reduce the
dynamics in the NiO spins and thus reduce the observed coercivity,
however, our data indicate the easy axis coercivity observed in
NiO/NiFe bilayers is primarily due to the static induced surface
anisotropy.  The ideal exchange anisotropy material for applications
would have a non-hysteretic hard axis loop and an easy-axis loop with
a large exchange bias and H$\rm _E >$ H$\rm _{ce}$.
		
Our observations on epitaxial oxide-AF/NiFe exchange-coupled bilayers
are consistent with those of Carey {\it et al.} who have done
extensive characterization of exchange couples using epitaxial NiO,
NiCoO films and NiO/CoO multilayers grown by reactive magnetron
sputtering.\cite{carey2} They report consistently smaller H$\rm _E$ in
epitaxial relative to polycrystalline bilayers deposited under similar
conditions.  They also observe exchange anisotropy in both (111) and
(001) oriented epitaxial bilayers, with consistently larger coercivity
in the (111) relative to the (001) oriented films.  Lai {\it et
al.}\cite{lai} report loop shifts in bilayers with epitaxial (001) and
(111) NiO films grown by metal-organic chemical vapor deposition
(MOCVD).  They observe unusually large and nearly isotropic coercivity
in both (001) and (111) oriented MOCVD based bilayer couples, however.

\section{Conclusions}

	We have shown that polycrystalline NiO/NiFe bilayers produce
larger loop shifts than epitaxial bilayers deposited simultaneously.
It appears that a larger surface anisotropy can be induced at a
polycrystalline relative to an epitaxial interface. The presence of
exchange anisotropy in (001)-oriented epitaxial NiO/NiFe layers shows
that an uncompensated interface is produced independent of the NiO
crystalline orientation. Our data show that in addition to the induced
surface unidirectional anisotropy, an induced cubic surface anisotropy
is needed to consistently model the hysteresis loops measured in
epitaxial NiO/NiFe bilayers.  Hysteresis loops of polycrystalline
bilayers are most accurately modeled by an induced surface
unidirectional plus uniaxial anisotropy.  The induced surface
anisotropies we observe are referenced only to the bias field applied
during deposition, and are independent of the NiO crystal structure.

\paragraph*{Acknowledgments}

Part of this work was performed under the auspices of the U. S.
Department of Energy (DOE) by LLNL under contract number
W-7405-ENG-48, and part was supported by DOE's Tailored
Microstructures in Hard Magnets Initiative. LEJ supported by DOE's
Science and Engineering Research Semester, and Partners in Industry
and Education programs.  We would like to thank Matt Carey and
Chih-Huang Lai for helpful discussions.  We thank Y. K. Kim and
Quantum Peripherals Colorado Inc. for their support during the early
stages of this project.  We thank Ron Musket for RBS measurements.

\clearpage

\begin{figure}
\caption{Hysteresis loops of the same polycrystalline
NiO500\AA/NiFe100\AA\ bilayer film are shown, one at a temperature
below the blocking temperature, T$\rm _b$, of NiO and one above T$\rm
_b$.  Above T$\rm _b$, the NiFe behaves as a free layer, magnetically
the same as a NiFe layer deposited on a non-magnetic substrate.  Below
T$\rm _b$ the interfacial exchange interaction induces a surface
anisotropy which shifts the NiFe loop away from the zero field axis
and raises its coercivity.}
\end{figure}

\begin{figure}
\caption{Comparison of x-ray spectra from NiO 500\AA/NiFe 100\AA\
bilayer films deposited simultaneously on polished single crystal MgO
(001) and amorphous substrates.  The bottom scan in a) shows that the
NiO on the oxidized silicon substrate is polycrystalline with a grain
size of approximately 150\AA.  The top scan shows the NiO (111) and
(022) Bragg peaks are absent in the bilayer grown on MgO.  The MgO
(002) substrate peak obscures the presence of the NiO (002) peak.  A
strong reflection is present from the NiFe (001) planes.  In b), phi
scans at the NiFe (022) and the MgO (022) Bragg angles are shown.  The
NiFe layer is epitaxially oriented relative to the MgO substrate,
confirming that the NiO layer is also epitaxial.}
\end{figure}

\begin{figure}
\caption{Magnetization data for two NiO 500\AA/NiFe 100\AA\ bilayer
films deposited simultaneously. In a) the easy-axis (H parallel to
H$\rm _b$, the bias field during growth) and hard-axis (H
perpendicular to H$\rm _b$) magnetization curves of the
polycrystalline bilayer couple is shown.  The easy-axis loop is
shifted by H$\rm _E$ = 52 Oe due to interfacial exchange anisotropy
with the NiO.  b) shows the same measurement as in a) for an epitaxial
(001) bilayer deposited on MgO.  The bias field, H$\rm _b$, applied
during deposition was aligned parallel to an in-plane MgO (100) axis.
The easy-axis loop is shifted by H$\rm _E$ = 20 Oe.  Discontinuities
in the hard-axis loop reveal the present of a cubic induced anisotropy
term that produces a local energy minimum parallel to the applied
field and perpendicular to H$\rm _b$.  In c) the transverse hard axis
magnetization, M$\rm _y$, for the polycrystalline (open circles) and
epitaxial (filled circles) are compared.  The smooth curve of the
polycrystalline M$\rm _y$ loop shows the magnetization vector rotates
continuously as the applied filed varies.  The plateau in the
epitaxial M$\rm _y$ loop confirms that the NiFe moment turns
discontinuously from a local energy well parallel to the applied field
to the deep unidirectional well perpendicular to it.}
\end{figure}

\begin{figure}
\caption{The same measurements as shown in figure 3 for a second set
of simultaneously deposited polycrystalline and epitaxial bilayers
except that here H$\rm _b$, the bias field during deposition, was
applied parallel to an in-plane (110) axis.  For the polycrystalline
films in a) the easy-axis loop is shifted by H$\rm _E$ = 66 Oe.  In b)
the epitaxial film has H$\rm _E$ = 36 Oe.  c) shows the transverse
hard axis magnetization loops.  The data are qualitatively similar to
those in figure 3, particularly the observation of discontinuities
indicating a cubic anisotropy with minima referenced to the bias field
axis.  Thus the cubic anisotropy is induced by the bias field and is
not influenced by the orientation of H$\rm _b$ relative to the NiO
crystal axis.}
\end{figure}

\begin{figure}
\caption{Easy axis magnetization for two bilayer films deposited in
reverse order simultaneously on MgO and an oxidized silicon substrate.
Both NiO layers grown on NiFe were polycrystalline.  The exchange bias
fields and coercivities of the two films are the same.  Differences in
the growth of NiO on NiFe compared to NiFe on NiO lead to the reduced
exchange anisotropy observed in these films.}
\end{figure}

\begin{figure}
\caption{The exchange anisotropy field, H$\rm _E$ for a wide variety
of NiO-500\AA/NiFe-t$\rm _{NiFe}$ films plotted vs. the easy-axis
coercivity, H$\rm _{ce}$.  The dotted line is a guide to the eye
indicating the relationship H$\rm _E$ = 1.8 H$\rm _{ce}$ + 14 Oe, so
the H$\rm _E$ to H$\rm _{ce}$ ratio has a limiting value of about 1.8.
This value appears to be a characteristic of the AF material since the
ratio observed for NiCoO/NiFe bilayers typically exceeds this value.
AF/NiFe bilayers using Mn-based antiferromagnetic layers greatly
exceed this value, giving much less coercivity per unit exchange
anisotropy shift.}
\end{figure}

\begin{figure}
\caption{Calculated easy axis (dashed) and hard axis (solid)
magnetization loops using an energy equation with a unidirectional and
cubic anisotropy (equation 5).  In a) we assume the magnetization
achieves the absolute minimum energy configuration. In b) we assume
the magnetization remains in local minima until the path to the
absolute minimum is unobstructed by an energy barrier.  The
calculations qualitatively reproduce the features observed in the
epitaxial NiO/NiFe bilayers.}
\end{figure}

\end{document}